\documentclass[twocolumn]{autart}

\newtheorem{theorem}{Theorem}

\newtheorem{lemma}{Lemma}

\newtheorem{remark}{Remark}

\newtheorem{assumption}{Assumption}

\usepackage{graphicx}
\usepackage{subfigure}
\usepackage{array}
\usepackage{color}
\usepackage{mathptmx}
\usepackage{amsmath}
\usepackage{amssymb}
\usepackage{cite}
\usepackage{enumerate} 
\usepackage{float}
\usepackage{natbib}
\usepackage{amsfonts} 
\usepackage{epsfig} 
\usepackage{epstopdf}
\usepackage{mathrsfs}
\usepackage{cases}

\pdfoutput=1

\begin{document}

\begin{frontmatter}

\title{Angle-Constrained Formation Control under Directed Non-Triangulated  Sensing Graphs (Extended Version)$^\star$}
\thanks{This work was supported in part by the National Key Research and Development Program of China under grant (No.2022YFB4701400/4701404) and by the National Natural Science Foundation of China under grant (62250710167, 61860206008, 61933012, 62203073), the Natural Science Foundation of Chongqing under Grant (CSTB2022NSCQ-MSX0577), and in part by the Central University Project under grant 2021CDJCGJ002. This paper is the extended version	of our paper published in Automatica. (Corresponding author: Gangshan Jing.)}

\author[LK1]{Kun Li}\ead{likun@cqu.edu.cn},
\author[LK1]{Zhixi Shen}\ead{shenzhixi@cqu.edu.cn},
\author[LK1]{Gangshan Jing}\ead{jinggangshan@cqu.edu.cn},
\author[LK1]{Yongduan Song}\ead{ydsong@cqu.edu.cn}

\address[LK1]{School of Automation, Chongqing University, Chongqing 400044, China.}


\begin{abstract}	
	Angle-constrained formation control has attracted much attention from control community due to the advantage that inter-edge angles are invariant under uniform translations, rotations, and scalings of the whole formation. However, almost all the existing angle-constrained formation control methods are limited to undirected triangulated sensing graphs. In this paper, we propose an angle-constrained formation control approach under a Leader-First Follower sensing architecture, where the sensing graph is \emph{directed} and \emph{non-triangulated}. Both shape stabilization and maneuver control are achieved under arbitrary initial configurations of the formation. During the formation process, the control input of each agent is based on relative positions from its neighbors measured in the local reference frame and wireless communications among agents are not required. \textcolor[rgb]{0,0,0}{We show that the proposed distributed formation controller ensures \emph{global exponential stability} of the desired formation for an $n$-agent system. Furthermore, it is interesting to see that the convergence rate of the whole formation is solely determined by \emph{partial} specific angles within the target formation.} The effectiveness of the proposed control algorithms is illustrated by carrying out experiments both in simulation environments and on real robotic platforms. 
\end{abstract}

\end{frontmatter}

\interdisplaylinepenalty=2500

\section{INTRODUCTION}\label{introduction}
The multi-agent system is a networked system composed of multiple agents that can interact with each other and implement controllers based on local information. Distributed control of multi-agent systems has gained significant attention in various categories such as consensus \cite{Song2018}, distributed optimization \cite{Song2020}, distributed localization \cite{Fangtac,Fangauto}, and formation control\cite{An2008}, etc. Among them, formation control is one of the most studied topics due to its wide applications in a variety of fields \cite{korea2015}, for example, reconnaissance of unmanned vehicles in extreme environments \cite{re}, coordination of mobile robots \cite{magnus2020}, and satellite formation flying \cite{sate}.

Generally speaking, distributed formation shape control is the problem of how to design a distributed controller based on available local information for a group of autonomous agents to form a specific formation shape \cite{jing2018}. The fundamental principle is that the formation shape can be uniquely determined by some local constraints in the graph. Depending on the type of local constraints, most of the existing formation control approaches can be categorized into displacement-based \cite{kang2015}, distance-based \cite{Anderson, park2015}, bearing-based \cite{zsy2016, LFF2019}, and mixed constraints \cite{kwon2020hybrid,fang2023integrated}. However, both displacement constraints and bearing constraints are dependent upon the global coordinate frame, which become difficult to utilize in specific scenarios such as the indoor and underwater environments. Distance-constrained formation control has gained significant attention since it only requires measurements in local coordinate frames. Nevertheless, it is not convenient to achieve scaling control for a distance-constrained formation. Moreover, most of existing distance-based formation control approaches only achieve local convergence.

In fact, inter-edge angle constraints are independent of the global coordinate frame and render the constrained formation the highest degree of freedom\footnote{Angles are invariant to motions including translation, rotation, and scaling, while relative displacement-based, distance-based, and bearing-based are only invariant to a subset of these motions.}, therefore are of our interest in this paper. In \cite{eren2003}, the authors first suggested an angle-based formation approach and presented several relevant problems. In \cite{basiri2010}, the authors studied the triangular formation problem with angle constraints and bearing-only measurements under undirected sensing graphs. In \cite{bishop2012}, triangular formation control based on mixed range and angle constraints was introduced. In \cite{magnus2021}, the authors presented infinitesimally shape-similar motions preserving invariance of angles and designed a decentralized heterogeneous formation control strategy for a class of triangulated frameworks. In \cite{Jing2019}, an angle-constrained shape determination approach (angle rigidity theory) was first proposed. With the aid of angle rigidity, the authors designed a distributed control law for formation stabilization based on angle constraints, and further obtained almost global convergence in \cite{Jing2020}. In \cite{CLM2020tac}, the authors presented a different angle rigidity theory by taking the sign of each angle into account. While requiring all angles to be defined in a common counterclockwise direction, the formation strategy in \cite{CLM2020tac} relaxed the measurements from relative positions to angles.

Unfortunately, all the angle-constrained approaches introduced above require the sensing graph to be undirected and triangulated, which usually cannot be satisfied in scenarios when each agent has a limited  sensing capability. In this paper, we achieve angle-constrained formation control in the two-dimensional space under a Leader-First Follower (LFF) structured sensing graph, which is directed and non-triangulated. LFF-based formation control has been extensively studied in the literature, e.g., \cite{ycb2009},\cite{ycb2011}, \cite{LFF2019}. Different from them, the formation controller proposed in this paper is based on angle constraints and ensures global convergence. To reflect the advantage of invariance of angles in translation, rotation, and scaling motions, we further study formation maneuver control. Both simulation and experimental tests are performed to support the theoretic analysis.

The main contribution of this paper is that we achieve angle-constrained formation control under non-triangulated sensing graphs for the first time, implying that our sensing graph condition is \textcolor[rgb]{0,0,0}{much weaker} than all the existing angle-based formation control references \cite{basiri2010, bishop2012, magnus2021, Jing2019, Jing2020, CLM2022tac}.


\emph{Notations:} In this paper, the notation used for the set of real numbers and $n$-dimensional Euclidean space are $\mathbb{R}$ and $\mathbb{R}^n$, respectively. Let $\textbf{1}_n$ be unit column vector with $n$-dimensional. $I_n$ represents the $n\times n$ identity matrix. $X^\top$ denotes the transpose of a matrix $X$. For a set of numbers $V$, $|V|$ is its cardinality. $\mathbb{N}^+$ represents the set of positive integers. $\mathcal{R}(\theta)=\begin{bmatrix}\textrm{cos}\theta&-\textrm{sin}\theta\\\textrm{sin}\theta&\textrm{cos}\theta\\ \end{bmatrix}$ is the 2-dimensional rotation matrix associated with rotation angle $\theta\in [0,2\pi)$. $||\cdot||$ is the Euclidean norm. $\det(A)$ denote the determinant of the square matrix $A$. $\triangle_{ijk}$ represents a triangle formed by three vertices $i,j,k$. $\textrm{O}(2)$ and $\textrm{SO}(2)$ are the orthogonal group and the special orthogonal group in $\mathbb{R}^2$ respectively. $\mathcal{O}(\cdot)$ is used to describe how closely a finite series approximates a given function.

\section{Problem Formulation}\label{II}
\subsection{Graph-Related Notions}
In this paper, a pair $\mathcal{G}=(\mathcal{V,E})$ is said to be a directed graph, where $\mathcal{V}=\{v_1,\dots,v_n \}$ is a vertex set corresponding to agents and $\mathcal{E}=\{(v_i,v_j)|i,j\in \mathcal{V},i\neq j\}$ is an edge set with pairs of directed edges. The ordered pair $(v_i,v_j)\in \mathcal{E}$ means an edge between $v_i$ and $v_j$ with an arrow directed from $v_i$ to $v_j$, i.e., vertex $v_i$ can access information from vertex $v_j$. Meanwhile, we say that $v_j$ is a neighbor of vertex $v_i$. The neighbor set of vertex $v_i$ is denoted by $\mathcal{N}_i=\{v_j\in \mathcal{V}|(v_i,v_j)\in \mathcal{E}\}$ and $|\mathcal{N}_i|$ is the cardinality of $\mathcal{N}_i$. For more details about directed graphs, we refer the readers to \cite{magnus2010}.

A pair $(\mathcal{G}, p)$ is said a framework, where $\mathcal{G}$ is a graph with $n$ vertices and $p = (p^\top_1,\dots, p^\top_n)^\top \in \mathbb{R}^{2n}$ is called a configuration, and $p_i \in \mathbb{R}^2$ is the coordinate in the global reference frame of vertex $i,~i = 1,\dots, n$. In this work, we use the framework $(\mathcal{G}_f,p^*)$ with $\mathcal{G}_f=(\mathcal{V},\mathcal{E}_f)$ to interpret the formation shape and $p^* = [{p^*_1}^\top,\dots, {p^*_n}^\top ]^\top$ is a configuration forming the desired shape. We use $\mathcal{G}_s= (\mathcal{V},\mathcal{E}_s)$ to interpret the sensing graph, which characterizes the sensing ability of agents. Given a configuration $p\in\mathbb{R}^{2n}$, we use the following set to specify the set of configurations that have the same shape as $p$:
\begin{multline}\label{te}
\mathscr{E}(p) \triangleq \{q\in \mathbb{R}^{2n}: q=c(I_n \otimes \mathcal{R}(\theta))p+\mathbf{1}_n\otimes\xi, \\ \mathcal{R}(\theta) \in \text{SO}(2), c\in \mathbb{R} \backslash \{0\}, \xi\in \mathbb{R}^2\}.
\end{multline}
where $c$ is the scale factor, $\theta$ is the rotation factor, and $\xi$ is the translation factor.

Given a framework $(\mathcal{G},p)$, a signed angle $\alpha_{ijk}$ represents the angle rotating from the vector $(p_i-p_j) $ to the vector $(p_k-p_j)$ under the counterclockwise direction. More specifically, $\alpha_{ijk}=\textrm{arccos}(b^\top_{ji}b_{jk})$ if sign$(\det(B_{ijk})) \le 0$, and $\alpha_{ijk}=2\pi-\textrm{arccos}(b^\top_{ji}b_{jk})$ otherwise, where $B_{ijk}=[b_{ji}, b_{jk}]$ and $b_{ji}=(p_i-p_j)/||p_i-p_j||$, respectively.

\subsection{Agent Dynamics and Sensing Capability}
Consider a group of $n$ agents modeled by a single integrator model:
\begin{equation}\label{p}
\dot{p}_i(t)=u_i(t),~i=1,\dots,n.
\end{equation}
where $p_i \in \mathbb{R}^2$ and $u_i \in \mathbb{R}^2$ are the position and the control input of agent $i$, respectively, with respect to the global coordinate system. Note that, in this paper, we consider the formation problem in a GPS-denied environment. In this scenario, different agents may have different local coordinate frames, each agent $i$ can only measure $p^i_j-p^i_i$ if $(i, j) \in  \mathcal{E}_s$, where $p^i_j$ denotes the position vector of agent $j$ in the local coordinate frame of agent $i$.

In this paper, we will utilize the \textcolor[rgb]{0,0,0}{minimally acyclic} LFF structure \cite{ycb2009} as the condition for the sensing graph $\mathcal{G}_s$, which is directed and non-triangulated. \textcolor[rgb]{0,0,0}{Designated agents $1$ and $2$ as the leader and the first follower, respectively.} Without loss of generality, we make the following assumption.
\begin{assumption}\label{as}
	\textcolor[rgb]{0,0,0}{The directed and non-triangulated sensing graph $\mathcal{G}_s$ is constructed such that: i) $|\mathcal{N}_1|$=0, $|\mathcal{N}_2|=1$, and $|\mathcal{N}_i|=2$, $\forall i \ge 3$; ii) If there is an edge between agents $i$ and $j$, where $i < j$, the edge must be $(j,i)$.}
\end{assumption}
\begin{assumption}\label{af}
	The target formation graph $\mathcal{G}_f=(\mathcal{V}, \mathcal{E}_f)$ contains $\mathcal{G}_s$ as a subgraph, $\mathcal{E}_f=\mathcal{E}_s\cup \{(j,k)\in \mathcal{N}^s_i,i\in\mathcal{V}\}$, and $(\mathcal{G}_f, p^*)$ is strongly nondegenerate\footnote{A framework in $\mathbb{R}^2$ is said to be strongly nondegenerate if two outgoing edges of the agent do not collinear.}, where $\mathcal{N}^s_i$ denotes the neighbor of agent $i$ in $\mathcal{G}_s$.
\end{assumption}
\begin{remark}
	\textcolor[rgb]{0,0,0}{Assumption \ref{as} indicates that $\mathcal{G}_s$ is a LFF type graph, which belongs to a class of acyclic minimally persistent graphs \cite{ycb2009}. This assumption is not restrictive as it only requires a minimum number of links for a framework to be rigid, , while rigidity has been commonly used as a condition for both sensing and formation graphs in many references, e.g., \cite{Jing2019,CLM2020tac}. Additionally, Assumption \ref{af} is widely used in the existing results for angle-constrained formation control. Note that we merely consider strongly non-degenerate formations in this paper. Compared with the references \cite{Lin2016,Lin2018} that require an additional generic assumption, Assumption \ref{af} is milder.}
\end{remark}
\begin{remark}
	\textcolor[rgb]{0,0,0}{We highlight that all the references on angle-constrained formation control require an undirected triangulated Laman graph as the sensing graph \cite{Jing2019,Jing2020,CLM2020tac}.} In this study, we consider angle-constrained formation control under sensing graphs with a LFF structure, which is a milder graph condition and has a significantly reduced number of edges compared with undirected triangulated Laman graphs. It should be pointed out that if $\mathcal{G}_s$ contains a LFF graph as a subgraph, our formation controller will be still valid since we can always ignore redundant sensing links. Note that in multi-agent coordination control, each edge between two agents usually represents an information flow. Hence, our approach benefits for reducing sensing burden.
\end{remark}

\subsection{Problem Statement}
Let $ \mathscr{E}(p^*)$ 
be a configuration manifold forming the shape of the target formation $(\mathcal{G}_f,p^*)$ with specified orientation factor $\theta=\theta^*$, scale factor $c=c^*$, and translation factor $\xi=\xi^*$, here $\mathscr{E}(\cdot)$ is defined in (\ref{te}). The first formation control problem in this paper aims to solve is as follows.


\emph{Problem 1:} (Shape Control)  Given target formation $(\mathcal{G}_f,p^*)$ and sensing graph $\mathcal{G}_s$, design a distributed control protocol $u_i$ for each agent $i$ with dynamics (\ref{p}) based on angle constraints in the target formation and relative position measurements $\{p^i_i-p^i_j,~(i,j)\in \mathcal{E}_s\}$ such that $p = [p^\top_1,\dots, p^\top_n ]^\top$ converges into $\mathscr{E}(p^*)$  asymptotically.



%

Maneuver control is a useful technique in practical formation control tasks. By appropriately adjusting the translation, rotation, and scale factors of the entire formation, a group of agents can dynamically respond to the complex environment during their motion. For example, a formation can be maneuvered to avoid obstacles, move through a narrow space and enclose specific objects. The formation maneuver control problem is formally stated below.

\emph{Problem 2:}  (Maneuver Control) Given piece-wise constant factors $c^*(t)$, $\theta^*(t)$, and $v_r^*(t)$ describing the target time-varying scale, orientation, and velocity of the formation $(\mathcal{G}_f,p^*)$, design a distributed control protocol $u_i$ for each agent $i$ with dynamics (\ref{p}) based on angle constraints in the target formation and relative position measurements $\{p^i_i-p^i_j,~(i,j)\in \mathcal{E}_s\}$ associated with the sensing graph such that $p$ converges to $\mathscr{E}(p^*)$ with $\theta(t)=\theta^*(t)$, $c(t)=c^*(t)$, and $\dot{p}_i(t)$ converges to $v_r^*(t)$ asymptotically. 

Note that, for ease of description, we drop the time parameter in the following of this paper, i.e., $x:= x(t)$, and relying on time will only be shown when introducing new concepts or symbols.

\section{Angle-Constrained LFF Formation Control}\label{III}
In this section, we propose distributed formation control laws under directed non-triangulated sensing graphs in the plane. The target formation will be characterized by constraints on angles. The agents are point agents, massless, and holonomic. The restriction on the sensing graph will be relaxed to directed non-triangulated graphs compared to the references \cite{basiri2010, bishop2012, magnus2021, Jing2019, Jing2020, CLM2022tac}.

\subsection{Angle Constraints in Target Framework}
Before the controller design, we will examine the Assumption \ref{af} and show how the angles in the target formation can be used to determine its shape. To specify the angle constraints in target formation, use $[k]$ to denote the $k$-th triangle corresponding to the vertex added with order $k$. Let $\mathcal{T}_{\mathcal{G}_f}=\{(i,j,k)\in \mathcal{V}^3:(k,i),(k,j)\in \mathcal{E}_f, i<j<k\}$, then the set of angles in the target formation to be exploited are denoted by $\mathcal{A}_{\mathcal{G}_f}=\{\dots,\alpha^*_{ikj},\alpha^*_{jki},\alpha^*_{kij},...: (i,j,k)\in \mathcal{T}_{\mathcal{G}_f}\}$, where $\alpha^*_{ikj}$, $\alpha^*_{jki}$, $\alpha^*_{kij}$ are the signed angles of $\triangle_{ijk}$	\textcolor[rgb]{0,0,0}{and $\alpha^*_{kij}$ is further termed as follower angle}.

In \cite[Lemma 2]{CLM2022auto}, the author showed that the shape of a non-degenerate triangle $\triangle_{ijk}$ can be uniquely determined by the following {\it linear constraint} based on signed interior angles:
\begin{equation}\label{clm}
A^{\triangle_{ijk}}_ip_i+A^{\triangle_{ijk}}_jp_j+A^{\triangle_{ijk}}_kp_k=0,
\end{equation}
where 
$A^{\triangle_{ijk}}_i=\left(\textrm{sin}\alpha_{jki}I_2-\textrm{sin}\alpha_{ijk}\mathcal{R}^\top(\alpha_{kij})\right) \in \mathbb{R}^{2\times2}$, $A^{\triangle_{ijk}}_j=\textrm{sin}\alpha_{ijk}\mathcal{R}^\top(\alpha_{kij})\in \mathbb{R}^{2\times2}$, $A^{\triangle_{ijk}}_k=-\textrm{sin}\alpha_{jki}I_2\in \mathbb{R}^{2\times2}$.

Based on this fact, we have the following lemma.

\begin{lemma}[Uniqueness of the target formation]\label{unique}
	Given a target formation $(\mathcal{G}_f,p^*)$ with $n$ vertices in $\mathbb{R}^2$ satisfying Assumption \ref{af}, the shape of $(\mathcal{G}_f,p^*)$ can be uniquely determined by angle constraints (\ref{clm}) in $(\mathcal{G}_f,p^*)$, i.e., $q\in\mathscr{E}(p^*)$ for $q=[q_1^\top,...,q_n^\top]^\top\in\mathbb{R}^{2n}$ if (\ref{clm}) holds for all $q_i$, $q_j$, $q_k$, $(i,j,k)\in\mathcal{T}_{\mathcal{G}_f}$.
\end{lemma}

\textcolor[rgb]{0,0,0}{{\it Proof:} For $(\mathcal{G}_f,p^*)$, we define the angle-constrained function corresponding to a given realizable angle constraint set $\mathcal{A}_{\mathcal{G}_f}$ as
	\begin{equation}
	f_{\mathcal{G}_f}(p^*)=(\dots,A^{[k]}_ip^*_i+A^{[k]}_jp^*_j+A^{[k]}_kp^*_k,\dots)^\top,
	\end{equation}
	where $(i,j,k)\in  \mathcal{T}_{\mathcal {G}_f}$, $A^{[k]}_{h}$, $h\in\{i,j,k\}$ is the angle-induced linear constraint elements in $\triangle_{ijk}$.}

\textcolor[rgb]{0,0,0}{Denote $p(n)$ as the configuration corresponding to $n$ vertices. Now, we prove the lemma by induction.}

\textcolor[rgb]{0,0,0}{For $n=3$, \cite[Lemma 2]{CLM2022auto} has shown that $f_{\mathcal{G}_f}(q(3))=0$ implies that all the three interior angles in the triangular are uniquely determined, i.e., $q(3)\in\mathscr{E}(p(3))$. Suppose that $q(n)\in\mathscr{E}(p(n))$ for $n=N$, we prove the case for $n=N+1$.}

\textcolor[rgb]{0,0,0}{Without loss of generality, suppose that $q_{N+1}$ is connected with $q_v$ and $q_w$, $v,w\in\{1,...,N\}$. Next we show $q_{N+1}$ can be uniquely determined by $q_v$, $q_w$, and an angle constraint.}

\textcolor[rgb]{0,0,0}{Considering (\ref{clm}), we have
	\begin{equation}
	A^{[N+1]}_{v}q_{v}+A^{[N+1]}_{w}q_{w}+A^{[N+1]}_{N+1}q_{N+1}=0,
	\end{equation}
	Since $\det\left(A^{[N+1]}_{N+1}\right)=\sin^2\alpha_{kij}>0$, which shows that the matrix $A^{[N+1]}_{N+1}$  is a non-singular matrix. }

\textcolor[rgb]{0,0,0}{Therefore,  
	\begin{equation}\label{q_{N+1}}
	q_{N+1}=-(A^{[N+1]}_{N+1})^{-1}\left(A^{[N+1]}_{v}q_{v}+A^{[N+1]}_{w}q_{w}\right).
	\end{equation}
	From the definition of $\mathscr{E}(p(N))$, we have $$q_v=c\mathcal{R}(\theta)p_v+\xi, ~~~~~q_w=c\mathcal{R}(\theta)p_w+\xi,$$ and $p_{N+1}=-(A^{[N+1]}_{N+1})^{-1}\left(A^{[N+1]}_{v}p_{v}+A^{[N+1]}_{w}p_{w}\right).$ 
	Then we have
	\begin{align}
	q_{N+1}=&-\left(A^{[N+1]}_{N+1}\right)^{-1}\left(A^{[N+1]}_{v}(c\mathcal{R}(\theta)p_v+\xi) \nonumber  \right. \\
	&\left. +A^{[N+1]}_{w}(c\mathcal{R}(\theta)p_w+\xi)\right) \nonumber \\
	=&-c\left(A^{[N+1]}_{N+1}\right)^{-1}\left(A^{[N+1]}_{v}\mathcal{R}(\theta)p_v +A^{[N+1]}_{w}\mathcal{R}(\theta)p_w\right)+\xi    \nonumber \\
	=&-c\left(A^{[N+1]}_{N+1}\right)^{-1}\left(\mathcal{R}(\theta)A^{[N+1]}_{v}p_v +\mathcal{R}(\theta)A^{[N+1]}_{w}p_w\right)+\xi    \nonumber \\	
	=&-c\mathcal{R}(\theta)\left[-(A^{[N+1]}_{N+1})^{-1}(A^{[N+1]}_{v}p_{v} +A^{[N+1]}_{w}p_{w}) \right] +\xi  \nonumber \\
	=&c\mathcal{R}(\theta)p_{N+1}+\xi.
	\end{align}
	where the third equality and the fourth equality follow from the fact that $A^{[N+1]}_{v}$, $A^{[N+1]}_{w}$, and $(A^{[N+1]}_{N+1})^{-1}$ are all scaled rotation matrices, and therefore are commutative with $\mathcal{R}(\theta)$.
	Thus, $q(N+1)\in\mathscr{E}(p(N+1))$. The proof is completed.} $\hfill\blacksquare$

\begin{remark}\label{sensing}
	Lemma \ref{unique} is based on signed interior angle constraints. In fact, when signed angles are utilized as sensing measurements, all the agents need to have a common understanding of the counter-clockwise direction. However, the signed angles used in this paper are constraints that can be calculated in advance by the target formation $(\mathcal{G}_f,p^*)$. Therefore, our distributed controller does not require the common counter-clockwise direction assumption.
\end{remark}

\subsection{Formation Stabilization}\label{fs}

Our goal is to solve Problem 1 via available sensing measurements and angle constraints in the target formation, under Assumptions \ref{as} and \ref{af}. Note that an angle always involves three agents. According to the LFF philosophy, the leader and the first follower are not able to meet any angle constraints based on their sensing capability. Therefore, we do not apply control to the first two agents, i.e., $u_i=0$ for $i=1,2$. The controllers for the rest of agents are designed as follows:
\begin{equation}\label{u}
u_k=-\left(A^{[k]}_k\right)^\top\left(A^{[k]}_ie_{ki}+A^{[k]}_je_{kj}\right),~ k=3,...,n,
\end{equation}
where $e_{ki}=p_i-p_k, e_{kj}=p_j-p_k$, $A^{[k]}_h$, $h\in \{i,j,k\}$ is the matrix in the angle-induced linear constraint (\ref{clm}) associated with $\triangle_{ijk}$, $\{i,j\in \mathcal{N}_k^s\}$ in $\mathcal{G}_f$, and the target formation $p^*$. \textcolor[rgb]{0,0,0}{The controller (\ref{u}) is linear since $A^{[k]}_i$ and $A^{[k]}_j$ are both constant matrices determined by angle constraints in the target formation.}
\begin{lemma}\label{distributed}
	The controller (\ref{u}) has the following properties:
	\begin{enumerate}
		\item The controller (\ref{u}) is a distributed strategy based on relative position measurements corresponding to sensing graph $\mathcal{G}_s$.
		\item The controller (\ref{u}) is independent of the global reference frame.
	\end{enumerate}
\end{lemma}

{\it Proof}: The first statement can be verified by observing the form of (\ref{u}) directly. Next we prove the second statement. Suppose the superscript $\{k\}$ indicates a quantity expressed in the local coordinate frame of the $k$-th agent. Let $Q^k_g \in \textrm{SO}(2)$ be the rotation matrix from the global frame to the $k$-th local frame. Then the local controller of agent $k$ can be represented as
\begin{align}
u^{\{k\}}_k=&~Q^k_gu_k=-Q^k_g\left(A^{[k]}_k\right)^\top\left(A^{[k]}_ie_{ki}+A^{[k]}_je_{kj}\right) \nonumber \\
=&-\left(A^{[k]}_k\right)^\top\left(A^{[k]}_ie^{\{k\}}_{ki}+A^{[k]}_je^{\{k\}}_{kj}\right)
\end{align}
where the third equality follow from the fact that $Q^k_gA^{[k]}_he_{kh}=A^{[k]}_hQ^k_ge_{kh}$, $h=i,j$. Thus the control law (\ref{u}) can be implemented in the local reference frame of each agent. $\hfill\blacksquare$

\begin{theorem}\label{T1}
	\textcolor[rgb]{0,0,0}{Under Assumptions \ref{as} and \ref{af}, consider an $n$-agent formation with dynamics (\ref{p}). By implementing the distributed control law (\ref{u}), the stacked vector of positions $p = [p^\top_1,\dots, p^\top_n ]^\top$ converges into $p^\dagger \in \mathcal{E}(p^*)$ exponentially, where
		\begin{equation}\label{pdagger}
		p^\dagger=c^\dagger(I_n\otimes\mathcal{R}(\theta^\dagger))p^*+\mathbf{1}_n\otimes\xi^\dagger.
		\end{equation}		
		Moreover, the convergence rate is solely determined by the follower angles $\alpha^*_{kij}$ within target framework, where $k=3,...,n$, and $i,j\in \mathcal{N}_k^s$.}
\end{theorem}
{\it Proof:} Note that $p_1(t)$ and $p_2(t)$ never change along the time, thus $p_1=p^\dagger_1, p_2=p^\dagger_2$. Moreover, according to (\ref{te}) and (\ref{pdagger}), $p^\dagger$ has the same shape as $p^*$, then the angle  linear constraints (\ref{clm}) induced from $p^*$ are still satisfied in $p^\dagger$. Let $\tilde{p}_k=p_k-p^\dagger_k$, $k=1,...,n$. In the following, we shall first establish the results for agent $3$ and then extend the proof for all $3 < k \le n$ by induction.

{\it Step 1~($k=3$):}
From (\ref{p}) and (\ref{u}), we have
\begin{align}\label{dp3}
\dot{\tilde{p}}_3=&-\left({A^{[3]}_3}\right)^\top\left(A^{[3]}_1e_{31}+A^{[3]}_2e_{32}\right).
\end{align}
Choose the Lyapunov function $V_1(t)=\frac{1}{2}\Vert\tilde{p}_3(t)\Vert^2$. The derivative of $V_1(t)$ along the trajectory of system (\ref{dp3}) is
\begin{align}\label{dv1}
\dot{V}_1=&-\tilde{p}_3^\top \left({A^{[3]}_3}\right)^\top\left(A^{[3]}_1e_{31}+A^{[3]}_2e_{32}\right) \nonumber \\
=&-\tilde{p}_3^\top \left({A^{[3]}_3}\right)^\top\left(A^{[3]}_1p^\dagger_1+A^{[3]}_2p^\dagger_2+A^{[3]}_3p_3\right) \nonumber \\
=&-\tilde{p}_3^\top \left({A^{[3]}_3}\right)^\top{A^{[3]}_3}\tilde{p}_3=-2 \sin\alpha^2_{312}V_1,
\end{align}
where the third equality and fourth equality follow from the fact that $A^{[3]}_1p^\dagger_1+A^{[3]}_2p^\dagger_2+A^{[3]}_3p^\dagger_3=0$ and $A^{[3]}_3=-\sin\alpha_{312}I_2\in \mathbb{R}^{2\times2}$ in (\ref{clm}). Solving the differential inequality (\ref{dv1}) on $[0, \infty)$, we have $V_1(t)= \exp(-2 \sin\alpha^2_{312}t)V_1(0)$. Thus, $\lim_{t\rightarrow \infty}V_1(t)=0$. That is, $p_3$ converges to $p^\dagger_3$ with an exponential rate $2 \sin\alpha^2_{312}$ over the time. More specifically, the equilibrium $p^\dagger_{3}$ is global exponential stable (GES).

\textcolor[rgb]{0,0,0}{{\it Induction Step:} Consider we generate the graph step-by-step in the analysis by adding a vertex $k~(k\ge4)$ with two outgoing edges to any two distinct vertices $i$ and $j$ of the previous graph, we obtain the following cascade system at each step:}
\begin{subequations}\label{model}
	\begin{align}
	\dot{\tilde{p}}_k=&f_k(\Pi_{k-1},\tilde{p}_k), \label{cacadesystem}\\
	\dot{\Pi}_{k-1}=&g_{k-1}(\Pi_{k-1}) \label{subsystem},
	\end{align}	
\end{subequations}
\textcolor[rgb]{0,0,0}{where $\Pi_{k-1}:=[\tilde{p}_3,\tilde{p}_4,...,\tilde{p}_{k-1}]$. Note that the GES of $\dot{\Pi}_{k-1}=0$ for (\ref{subsystem}) was already established in the previous step. Therefore, we only need to check if (\ref{cacadesystem}) is input-to-state stable (ISS) with respect to input $\Pi_{k-1}$.}

According to (\ref{u}), the dynamics of agent $\tilde{p}_k$ is
\begin{align}\label{pk}
\dot{\tilde{p}}_k=&-\left({A^{[k]}_k}\right)^\top\left(A^{[k]}_ie_{ki}+A^{[k]}_je_{kj}\right) \nonumber \\
=&-\left({A^{[k]}_k}\right)^\top\left(A^{[k]}_i\tilde{p}_{i}+A^{[k]}_j\tilde{p}_{j}+A^{[k]}_k\tilde{p}_k\right),
\end{align}
where $\{i,j\in \mathcal{N}^s_k\}$.
We consider (\ref{pk}) as a cascade system with $\tilde{p}_{i}$ and $\tilde{p}_{j}$ being inputs to the unforced system
\begin{equation}\label{unforcepk}
\dot{\tilde{p}}_k=f_k(0,0,\tilde{p}_k)=-\left({A^{[k]}_k}\right)^\top A^{[k]}_k\tilde{p}_k=-\sin\alpha^2_{kij}I_2\tilde{p}_k.
\end{equation}
\textcolor[rgb]{0,0,0}{Likewise, the unforced error system (\ref{unforcepk}) is GES at $p^\dagger_k$. As a result, (\ref{cacadesystem}) is ISS by \cite{khalil1996nonlinear}. Finally, we can conclude that $[\Pi_{k-1},\tilde{p}_k]=0$ in (\ref{model}) is GES \cite{khalil1996nonlinear}. Repeating this process until $k = n$ leads to the conclusion that $\tilde{p}=[\tilde{p}^\top_1,...,\tilde{p}^\top_n]^\top=0$ is GES, which implies $\lim_{t\rightarrow\infty}p_k=p^\dagger_k$, $k=3,...,n$.}

Next we derive the detailed expressions of $c^\dagger$, $\theta^\dagger$, and $\xi^\dagger$. Note that 
\begin{subnumcases}{}
p^\dagger_1=c^\dagger\mathcal{R}(\theta^\dagger)p^*_1+\xi^\dagger, \label{p1}\\ 
p^\dagger_2=c^\dagger\mathcal{R}(\theta^\dagger)p^*_2+\xi^\dagger.
\end{subnumcases}
According to $p_1^\dagger=p_1(0)$ and $p_2^\dagger=p_2(0)$, we have
\begin{equation}\label{eq p10-p20}
p_1(0)-p_2(0)=c^\dagger\mathcal{R}(\theta^\dagger)(p^*_1-p^*_2).
\end{equation}
Since the rotation matrix $\mathcal{R}(\theta^\dagger)$ multiplication has no effect on the vector size,
\begin{align}
||p_1(0)-p_2(0)||=&c^\dagger||\mathcal{R}(\theta^\dagger)(p^*_1-p^*_2)|| \nonumber \\
=&c^\dagger||(p^*_1-p^*_2)||.
\end{align}
Therefore, 
\begin{equation}\label{cdager}
c^\dagger=\frac{\|p_1(0)-p_2(0)\|}{\|p_1^*-p_2^*\|}.
\end{equation}
Invoking (\ref{cdager}) to (\ref{eq p10-p20}), therefore
\begin{equation}\label{theta}
\theta^\dagger=\begin{cases}
\textrm{arccos}\left(b^\top_{21}b^*_{21}\right),~\textrm{if}~b^\top_{21}R(\frac{\pi}{2})b^*_{21}\ge 0,   \\[2ex] 
2\pi-\textrm{arccos}\left(b^\top_{21}b^*_{21}\right), \textrm{otherwise}.
\end{cases}
\end{equation}
where $b_{21}=\frac {p_1(0)-p_2(0)}{||p_1(0)-p_2(0)||}, b^*_{21}=\frac{p^*_1-p^*_2}{||p^*_1-p^*_2||}.$

Substituting (\ref{theta}) into (\ref{p1}), yields:
\begin{equation}
\xi^\dagger=p_1(0)-c^\dagger R(\theta^\dagger)p^*_1.
\end{equation}
The proof is completed. $\hfill\blacksquare$

There is an implicit assumption in Theorem \ref{T1} that $p_1(0)\neq p_2(0)$. Since the leader and the first follower keep stationary, we assume that the first two agents do not coincide initially. There are no additional limitations for the initial positions of the rest of agents.

\begin{remark}
	In contrast to distance-based methods, the angle-constrained approach  is more convenient to achieve scaling control owing to the advantage of invariance of angles in translation, rotation, and scaling motions \textcolor[rgb]{0,0,0}{(for more information see Section \ref{maneuvercontrol}).} Moreover, we obtain global exponential convergence in this paper, providing better performance and greater robustness to nonlinearity, perturbations, etc. Therefore, it is convenient to apply (\ref{u}) to real robotic platforms, as will be shown in Section \ref{IV}. In particular, it is interestingly noticed from (\ref{dv1}) that the convergence rate of the formation is only determined by the follower angles $\{\sin_{kij}\}_{(i,j\in\mathcal{N}^s_k)}$ obtained by the target formation beforehand. Therefore, the convergence rate of the formation algorithm can reaches maximum when each follower agent in the target formation maintains an angle of either $\pi/2$ or $3\pi/2$ with its neighboring agents. The proposed controllers can also be extended to 3-D space based on the novel 3-D angle-based constraint \cite{Fangtac}.
\end{remark}
\begin{remark}
	As an inspiration, here we propose a single-integrator-based control law. More elaborate control laws may be required in practice to meet more complicated dynamics. In fact, our reasoning remains valid for such control laws provided the notion of equilibrium is appropriately redefined based on connections between different dynamic models. For example, in order to control the differential-drive robots to achieve specific formation, single-integrator dynamics can always be mapped to unicycle models through a near-identity diffeomorphism (NID) \cite{magnus2020}.
\end{remark}

Avoiding collisions among agents is an important issue in practical formation control \cite{Fangauto}. In fact, an algebraic condition for collision avoidance among triangles in $\mathcal{G}_f$ can be derived if the control law (\ref{u}) is implemented for each agent $i$ after its two neighboring agents are stabilized, see the following lemma.
\begin{lemma}[Collision-free between agent $i$ and its neighbors]\label{collision-free}
	Suppose that the control law (\ref{u}) for each agent $i\geq3$ is implemented after its two neighboring agents are stabilized. Agent $i$ never collides with its neighboring agents $j, k$ if 
	\begin{equation}\label{c3}
	||p_i(0)-p^*_i||<\textrm{min}\{||p^*_i-p^*_j||,||p^*_i-p^*_k||\}.
	\end{equation}
\end{lemma}
{\it Proof:} Notice that agents $p_j$ and $p_k$ have been stabilized to $p^*_j$ and $p^*_k$. Therefore, agent $i$ never collides with agent $j$ if $||p_i-p_j||=||p_i-p^*_j||>0,~\forall t\ge0$. By Theorem \ref{T1}, we know $||p_i-p^*_i|| \le ||p_i(0)-p^*_i||$. It follows that
\begin{align}
||p_i-p^*_j||=&||(p_i-p^*_i)+(p^*_i-p^*_j)|| \nonumber \\
\ge &||p^*_i-p^*_j||-||p_i-p^*_i||  \nonumber \\
\ge &||p^*_i-p^*_j||-||p_i(0)-p^*_i||.
\end{align}
Thus, the collision between agents $i$ and $j$ can be avoided once $||p_i(0)-p^*_i||<||p^*_i-p^*_j||$. Similarly, collision-free between agent $i$ and $k$ can be achieved if $||p_i(0)-p^*_i||<||p^*_i-p^*_k||$.
Therefore, condition (\ref{c3}) guarantees the collision-free between agent $i$ and its neighbors.  $\hfill\blacksquare$

\subsection{Formation Maneuver Control}\label{maneuvercontrol}
Problem 2 requires the agents to not only stabilize a target shape asymptotically but also move with a common velocity eventually with desired translation, rotation and scaling factors. To achieve this goal, we show that controlling only partial agents is sufficient for maneuvering the whole formation.

In fact, according to (\ref{cdager})-(\ref{theta}) in Theorem \ref{te}, we observe that the scale, orientation and translation of the whole formation only depend on some local constraints between the leader and the first follower. We summarize the details in the following lemma.

\begin{lemma}\label{le translation orientation scale}
	Given a target formation $(\mathcal{G}_f,p^*)$ satisfying Assumption \ref{af}, the following statements hold:
	
	(i) The bearing between the leader and the first follower determines the orientation of the target formation.
	
	(ii) The distance between the leader and the first follower determines the scale of target formation.
	
	(iii) Upon fixing the relative position between the leader and the first follower, the translation of the leader determines the translation of the target formation.
\end{lemma}

Recall Problem 2, when designing the maneuver control law, we have factors $c^*(t)$, $\theta^*(t)$, $v^*_r(t)$ and $p^*$ at hand. Owing to Lemma \ref{le translation orientation scale}, we only need to constrain the relative position between the leader and the first follower as $\delta^*_{12}(t)$. Let $\{t_0,t_1,\dots,t_{\sigma} \}_{\sigma \in \mathbb{N}^+}$ with $t_0=0$ be the time instants that $v^*_r(t)$ and $\delta^*_{12}(t)$ switch values such that the target translation, orientation, scale, and velocity can be achieved. More specifically, $v^*_r(t)=v_\tau,\delta^*_{12}(t)=\delta_\tau$ for $t\in [t_{\tau-1},t_\tau)$, $\tau=1,...,\sigma$. 


Now, we design the maneuver control algorithm for all the agents as
\begin{align}\label{man}
&\dot{p}_1=v^*_r(t), \nonumber \\
&\dot{p}_2=v^*_r(t)-\left(e_{12}-\delta^*_{12}(t)\right), \\
&\dot{p}_k=v^*_r(t)-\left(A^{[k]}_k\right)^\top\left(A^{[k]}_ie_{ki}+A^{[k]}_je_{kj}\right), \nonumber
\end{align}
where $ i,j\in \mathcal{N}^s_k$, $3\le k\le n$, $e_{ki}=p_{i}-p_k,e_{kj}=p_{j}-p_k$, $A^{[k]}_{h}$, $h\in\{i,j,k\}$ is the matrix in the angle-induced linear constraint (\ref{clm}) associated with $\triangle_{ijk}$, $\{j,k\in \mathcal{N}_i^s\}$ in $\mathcal{G}_f$ and the target configuration $p^*$.

Note that each agent requires the common knowledge $v^*_r(t)$ when implementing (\ref{man}). To achieve this in a distributed manner, an approach is to make the agents communicate with each other via a communication graph. Suppose that only the leader knows $v^*_r(t)$. Then, all the agents can eventually obtain the reference velocity information via communications as long as the communication graph has a spanning tree with the leader as the root. In addition, motivated by the novel complex-laplacian-based algorithm designed by \cite{fang2023distributed}, a potential method that utilizes signed angle measurements can be developed and will be a topic of our future work.

\begin{remark}
	In real applications, one may only focus on controlling the scale or the orientation of the whole formation during the maneuver control. According to Lemma \ref{le translation orientation scale}, we only need to constrain the relative distance or the relative bearing between the leader and the first follower. To this end, a distanced-based or a bearing-based controller for the first follower can be designed as
	\begin{equation}\label{distance}
	u_2=-\left(||e_{12}||^2-||\delta^{*}_{12}||^2\right)e_{12},
	\end{equation}
	or
	\begin{equation}\label{bearing}
	u_2=P_{b_{12}}b^*_{12},
	\end{equation}
	where $P_{b_{12}}=I_2-b_{12}b^\top_{12}$ and $b^*_{12}=\frac{\delta^*_{12}}{||\delta^*_{12}||}$. 
	
	The effectiveness of the distance-based control law (\ref{distance}) and bearing-based control law (\ref{bearing}) have been proved in References \cite[Theorem 3.1]{park2015} and \cite[Lemma 4]{LFF2019}, respectively, interested readers may refer to these references.
\end{remark}

\begin{theorem}\label{T2}
	Consider an $n$-agent formation with dynamics (\ref{p}). By implementing the distributed maneuver control law (\ref{man}), for each time interval $t\in [t_{\tau-1},t_\tau)$, $\tau=1,...,\sigma$, we have
	\begin{align}
	&||p_2(t)-p_1(t)-\delta^*_{12}(t_{\tau-1})|| = \mathcal{O}\left(e^{-(t-t_{\tau-1})}\right), \label{mane1} \\
	&||p(t)-p^\ddagger||=\mathcal{O}\left(e^{-c_1(t-t_{\tau-1})}\right),~p^\ddagger\in \mathscr{E}(p^*), \label{shape} \\
	&||\dot{p}_k(t)-v^*_r(t_{\tau-1})|| =\mathcal{O}\left(e^{-c_2(t-t_{\tau-1})}\right), \label{mane2}
	\end{align}
	where $c_1, c_2 >0, k=3,...,n$.
\end{theorem}
{\it Proof:} The validity of (\ref{mane1}) can be proved by constructing $V_{e_1}=||p_2-p_1-\delta^*_{12}||^2$. The proof of (\ref{shape}) and (\ref{mane2}) can be obtained by a similar approach to Induction Step in proof of Theorem \ref{T1}. Thus the proof is omitted here. $\hfill\blacksquare$

\section{EXPERIMENTS}\label{IV}
\begin{figure}[htbp]	
	\centering
	\subfigure[]{
		\begin{minipage}[t]{0.43\linewidth}
			\centering
			\includegraphics[width=1.5in]{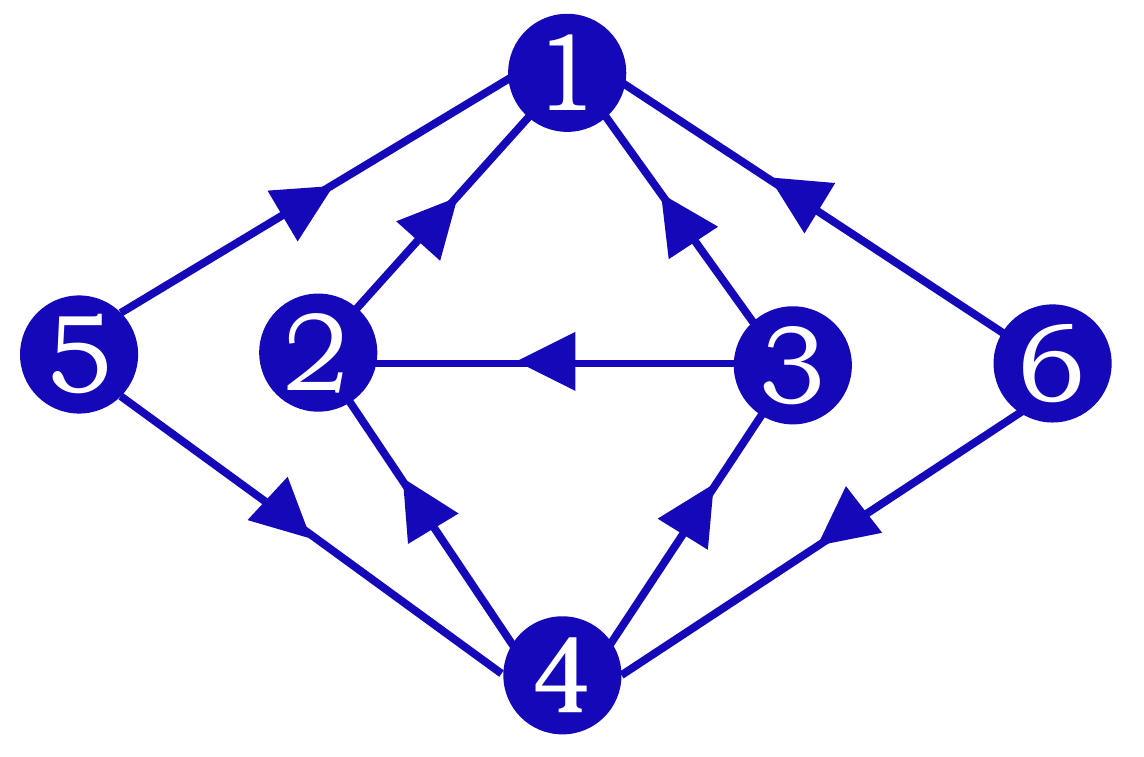}\label{sensinggraph}
		\end{minipage}
	}		
	\subfigure[]{
		\begin{minipage}[t]{0.43\linewidth}
			\centering
			\includegraphics[width=1.5in]{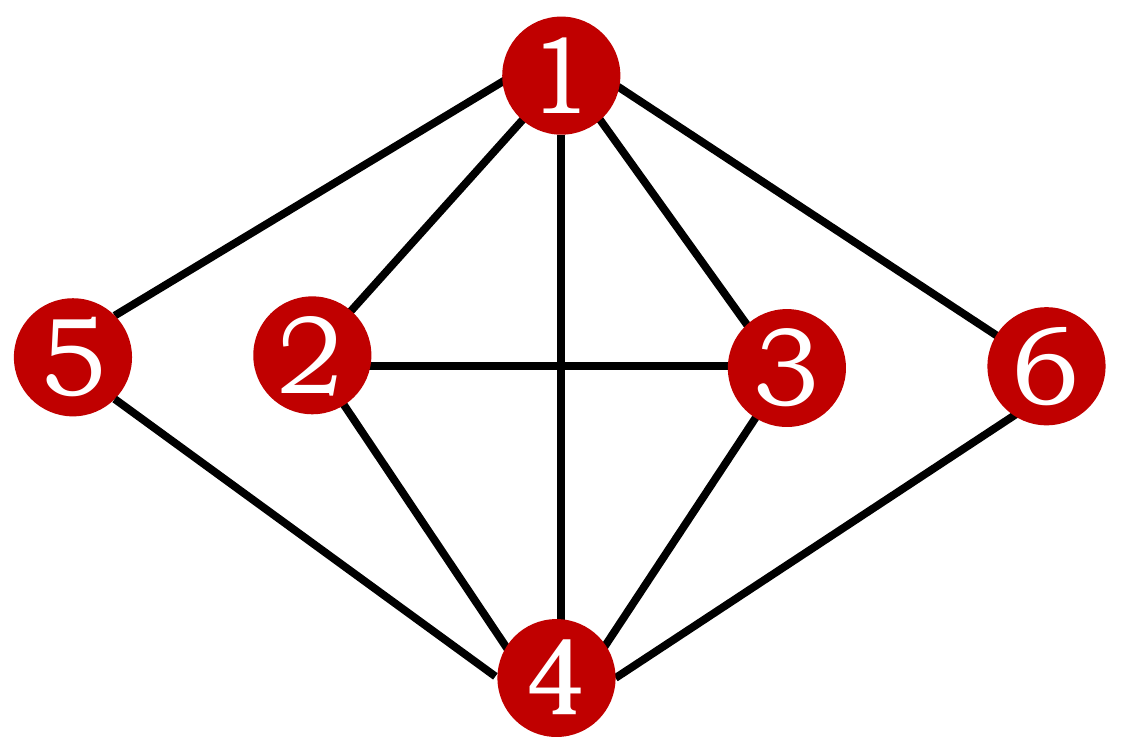}\label{targetgraph}
		\end{minipage}
	}
	\centering
	\caption{The formation graph. (a) the sensing graph $\mathcal{G}_s$; (b) the target formation ($\mathcal{G}_f,p^*$).}
\end{figure}

In this section, we validate the effectiveness of Theorems \ref{T1} and \ref{T2} using a team of differential-drive robots via the ``Robotarium" platform\footnote{An illustrative simulation video is available at: https://youtu.be/LUcyMYfR7y8.}. Each robot is of 11 cm wide, 10 cm long, 7cm tall, and has a maximum speed 20 cm/s linearly and a maximum rotational speed 3.6 rad/s (about 1/2 rotation per second). For more details about Robotarium, please refer to \cite{magnus2020}. Consider  the directed non-triangulated sensing graph and the target formation ($\mathcal{G}_f,p^*$) are shown in Fig. \ref{sensinggraph} and Fig. \ref{targetgraph} respectively satisfying Assumptions \ref{as} and \ref{af}. The set of angles in the target formation to be exploited is given as $\mathcal{A}_{\mathcal{G}_f}=\{\alpha^*_{123}=90^\circ,\alpha^*_{213}=315^\circ,\alpha^*_{324}=45^\circ,\alpha^*_{423}=270^\circ,\alpha^*_{415}=63.43^\circ,\alpha^*_{514}=306.87^\circ,\alpha^*_{416}=296.57^\circ,\alpha^*_{614}=53.13^\circ\}$.
\subsection{Formation Shape Stabilization} 
\begin{figure}[htbp]	
	\centering
	\subfigure[]{
		\begin{minipage}[t]{0.4\linewidth}
			\includegraphics[width=1.5in]{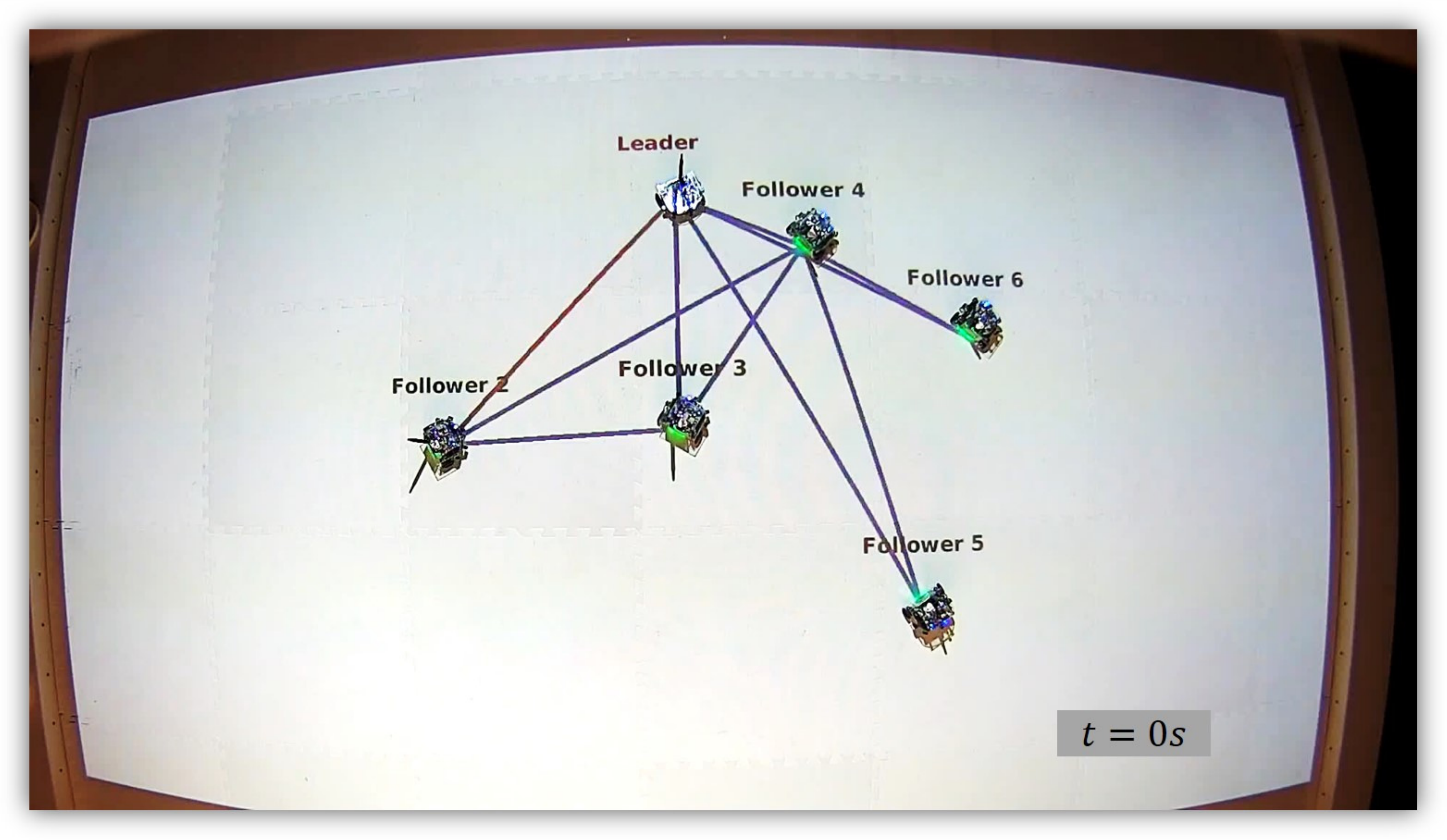}\label{shape a}
		\end{minipage}
	}
	\subfigure[]{
		\begin{minipage}[t]{0.4\linewidth}
			\includegraphics[width=1.5in]{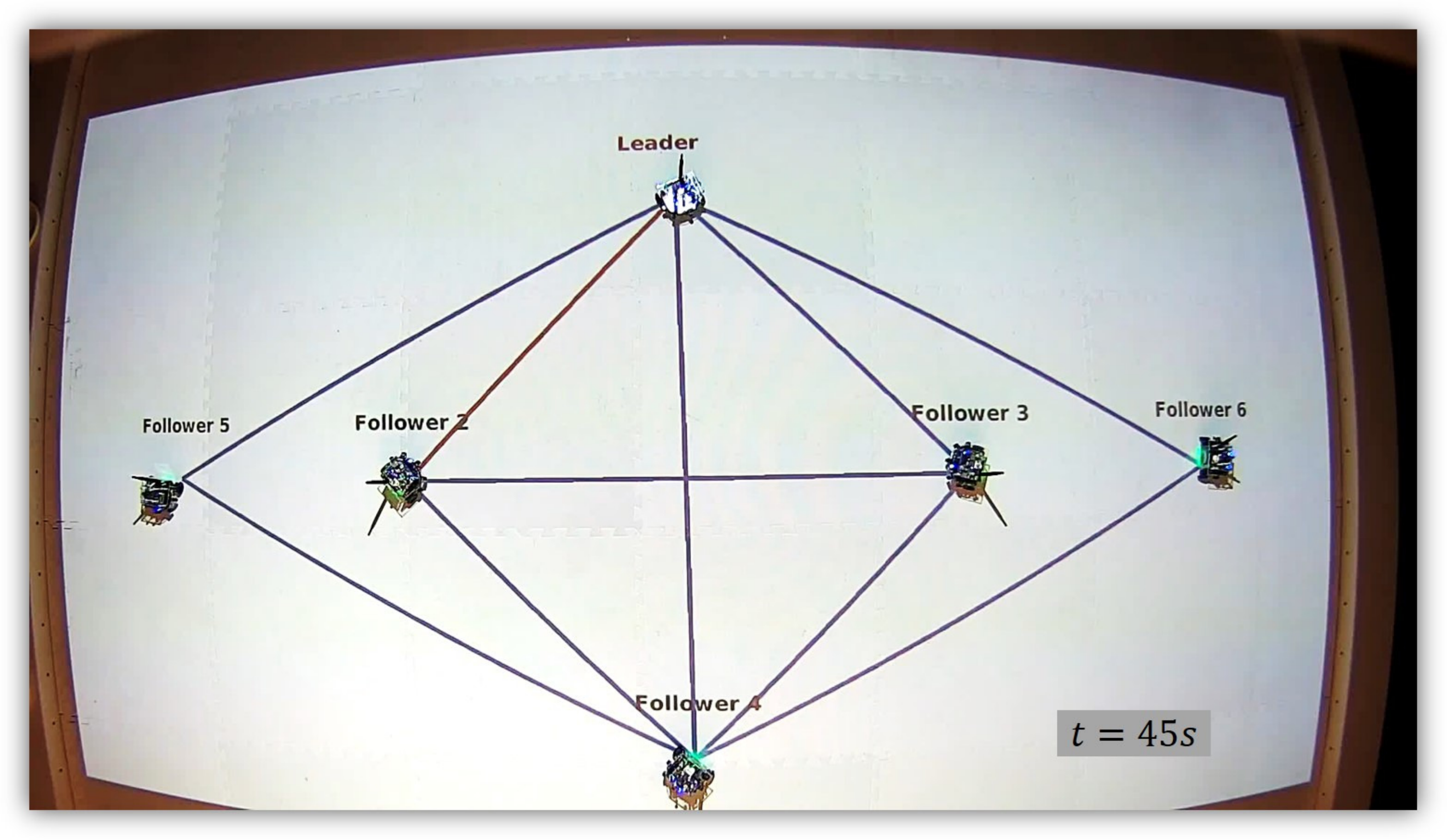}\label{shape b}
		\end{minipage}
	}
	\subfigure[]{
		\begin{minipage}[t]{1\linewidth}
			\centering	
			\includegraphics[width=2.8in]{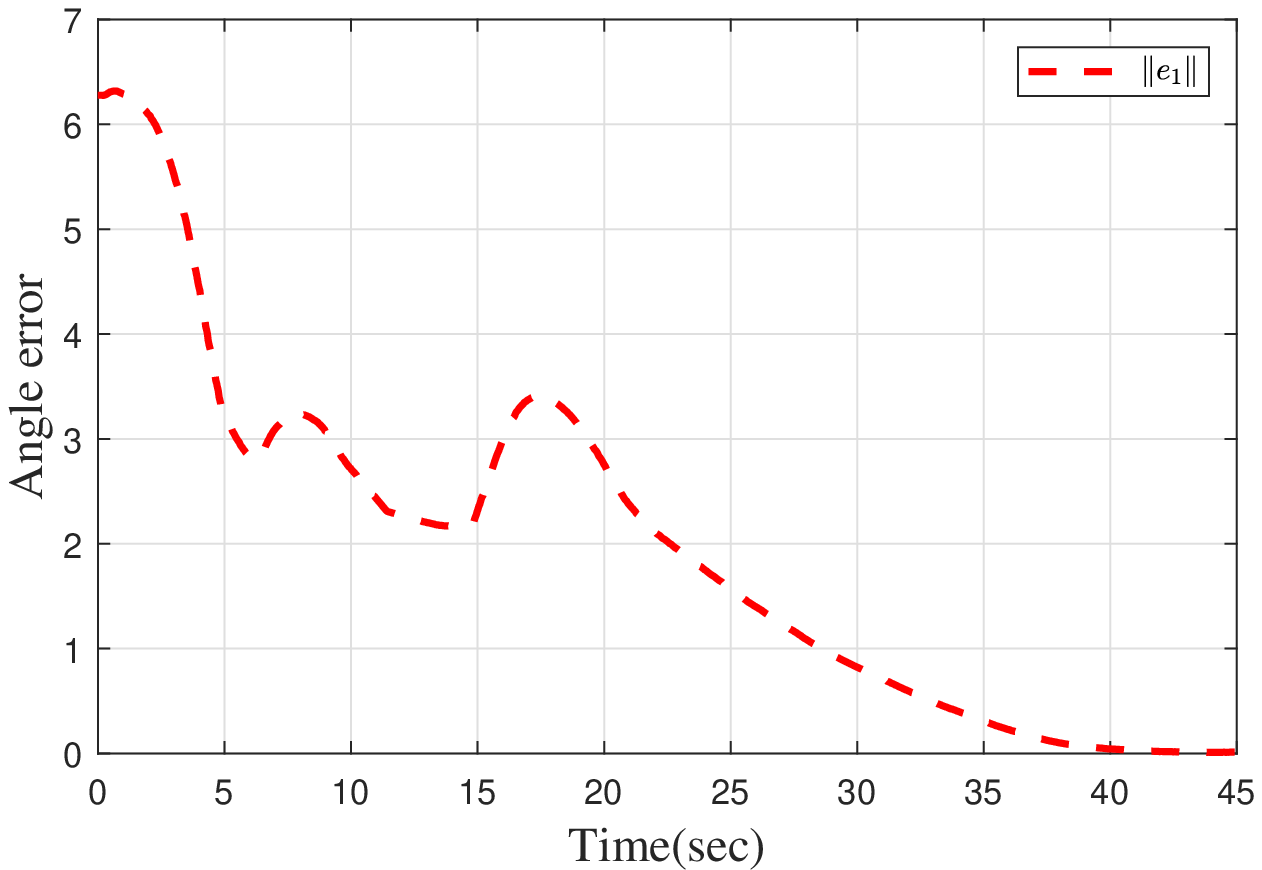}\\	\label{shape c}
	\end{minipage}}
	\centering
	\caption{The formation shape control law (\ref{u}) was implemented on a team of six differential drive robots. (a) the target formation ($\mathcal{G}_f,p^*$); (b) the sensing graph $\mathcal{G}_s$; (c) initial positions of six robots on the Robotarium; (d) the six robots eventually form the target formation shape; (e) the angle errors $\|e_1\|=\sum_{\alpha_{ijk} \in \mathcal{A}_{\mathcal{G}_f}}\|\alpha_{ijk}-\alpha^*_{ijk}\|$ converge to zeros asymptotically.}
\end{figure}

To fit the $12\times14$-ft$^2$ testbed, the initial positions of six differential-drive robots are chosen as $p(0)=[0, -0.5, 0, 0.25, \\ 0.5, 0.6; 0.5, 0, 0.025, 0.4,-0.35, 0.2]$ for better visual presentation, which are shown in Fig. \ref{shape a}. The posture vectors are randomly chosen in the two-dimensional space. By implementing the formation shape control law (\ref{u}), the eventual positions of six differential-drive robots and the corresponding evolution of angle errors in $\mathcal{A}_{\mathcal{G}_f}$ are presented respectively in Fig. \ref{shape b} and Fig. \ref{shape c}, showing that the desired formation shape is asymptotically achieved.

\subsection{Formation Maneuver Control}
\begin{figure}
	\subfigure[]{
		\begin{minipage}[t]{0.41\linewidth}
			\centering
			\includegraphics[width=1.5in]{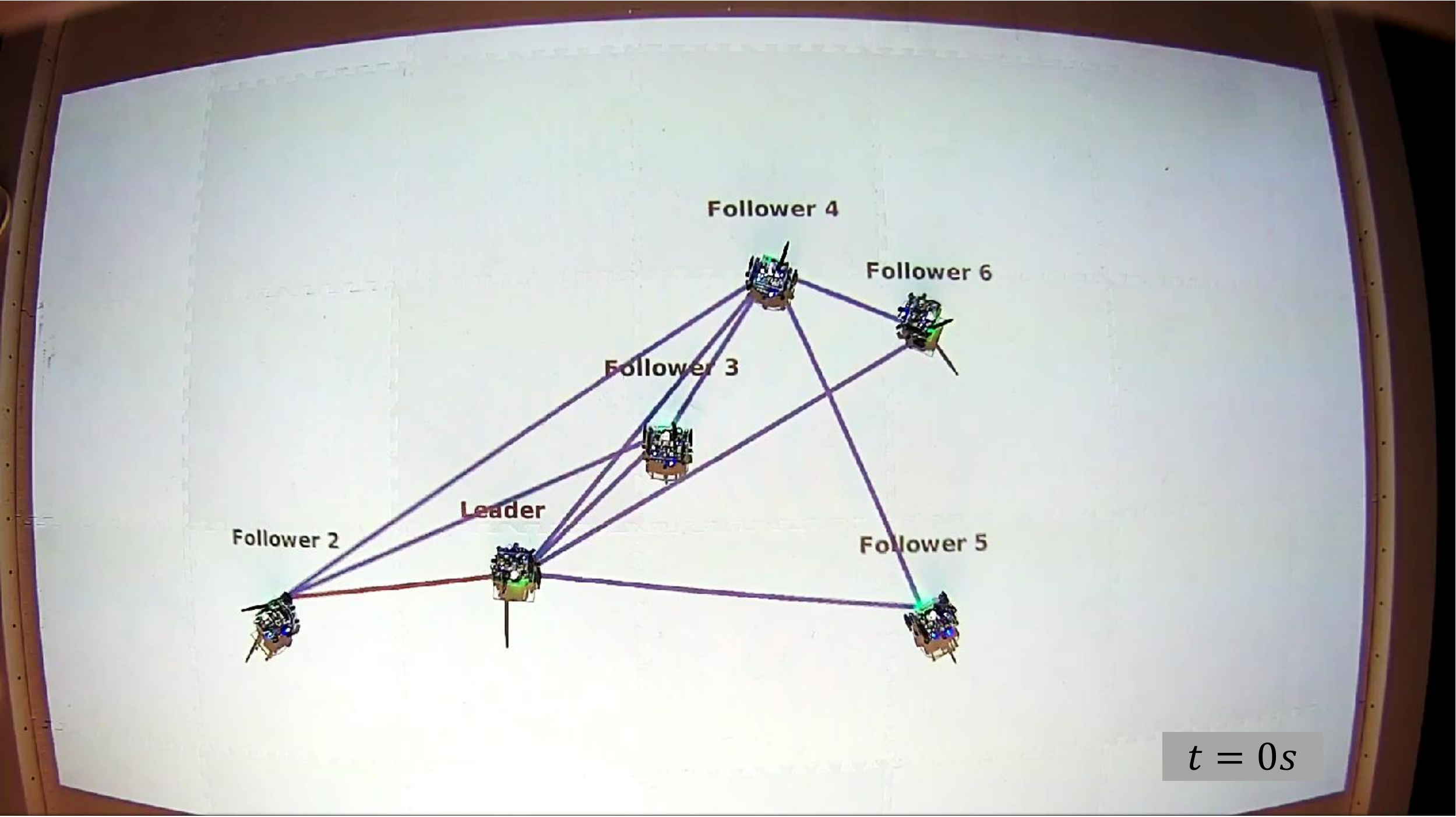}\label{m1}
		\end{minipage}
	}
	\subfigure[]{
		\begin{minipage}[t]{0.41\linewidth}
			\centering
			\includegraphics[width=1.5in]{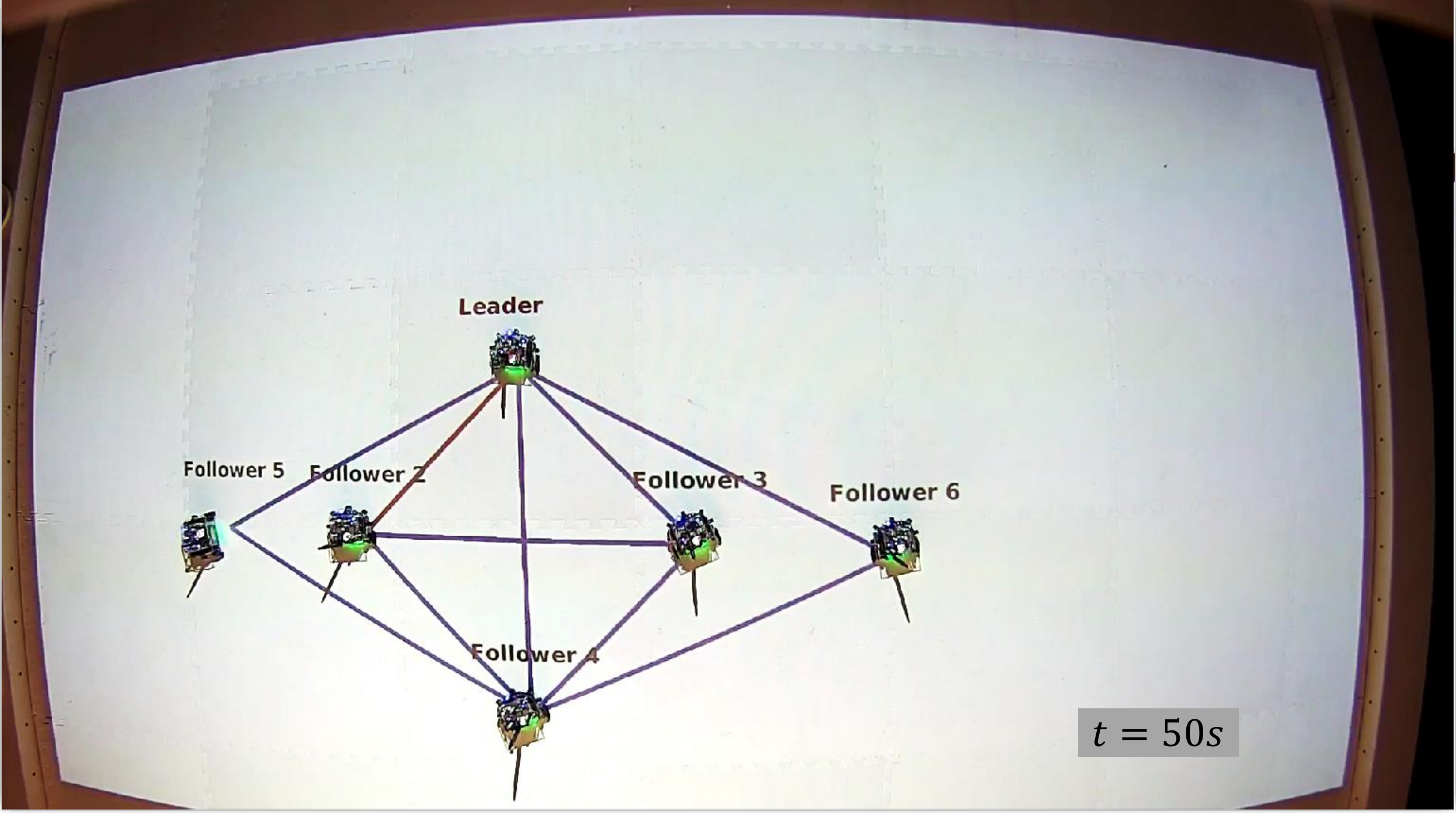}
		\end{minipage}
	}
	\subfigure[]{
		\begin{minipage}[t]{0.41\linewidth}
			\centering
			\includegraphics[width=1.5in]{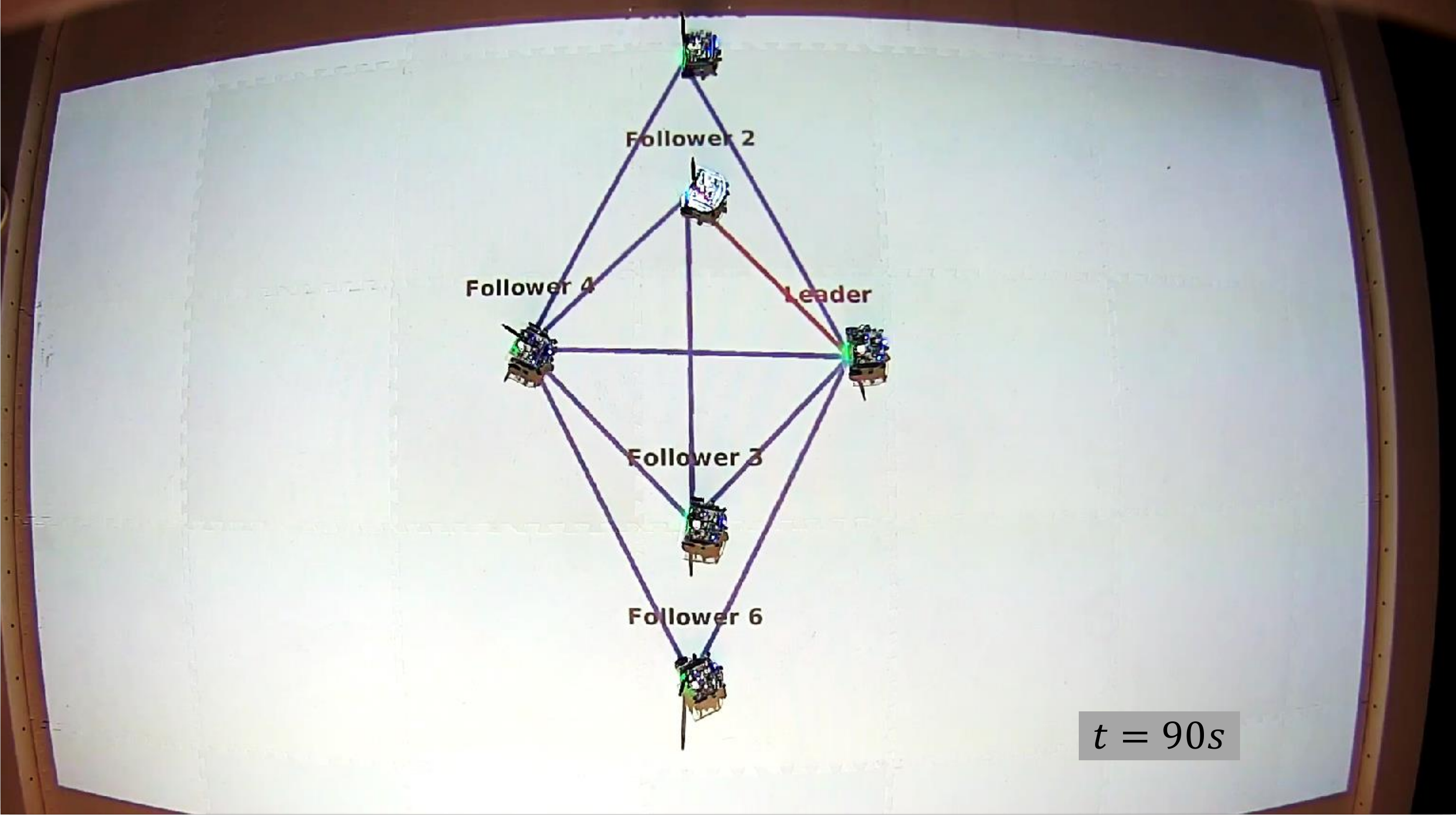}
		\end{minipage}
	}
	\subfigure[]{
		\begin{minipage}[t]{0.41\linewidth}
			\centering
			\includegraphics[width=1.5in]{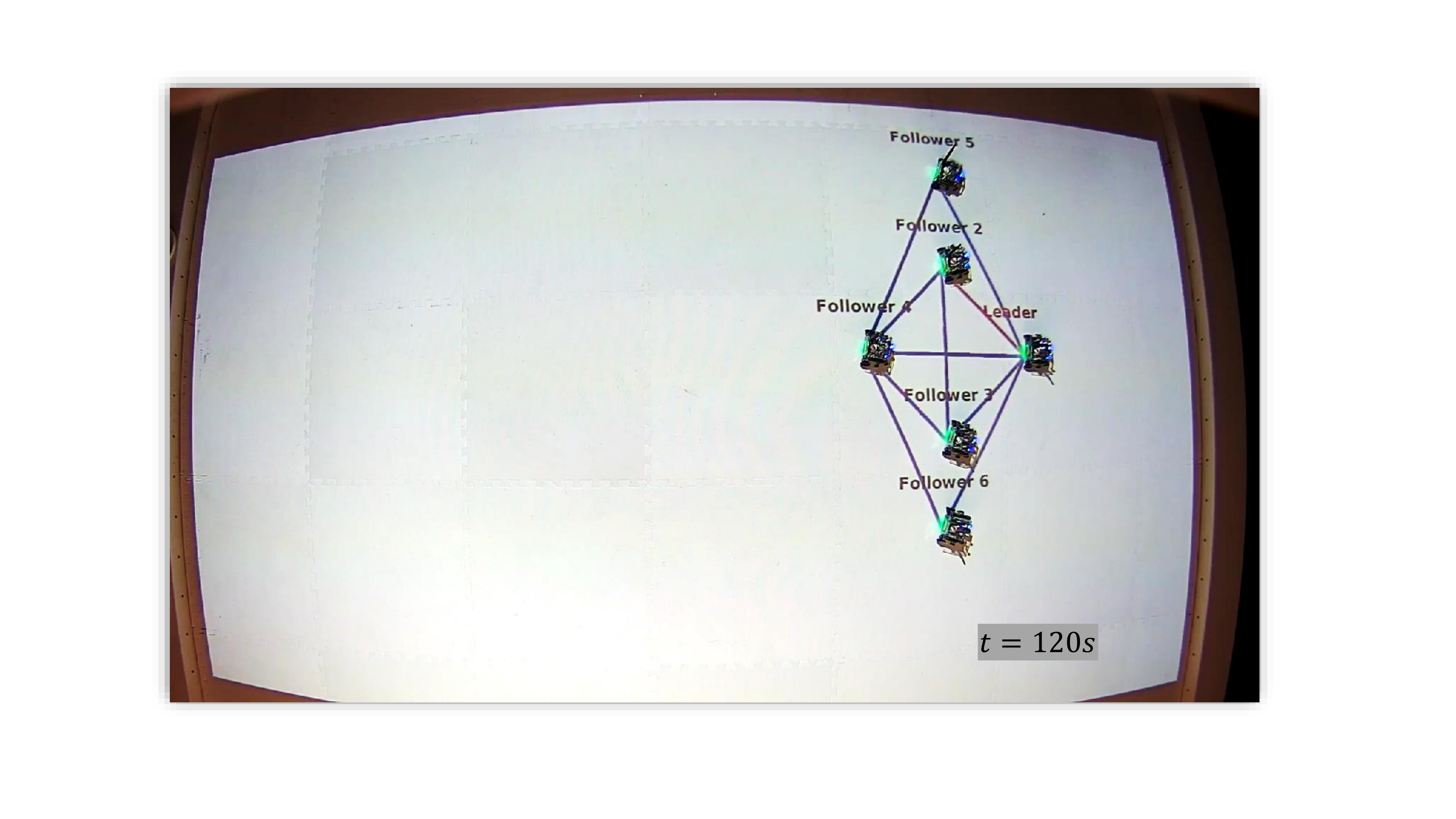}
		\end{minipage}
	}
	\subfigure[]{
		\begin{minipage}[t]{1\linewidth}
			\centering	
			\includegraphics[width=2.8in]{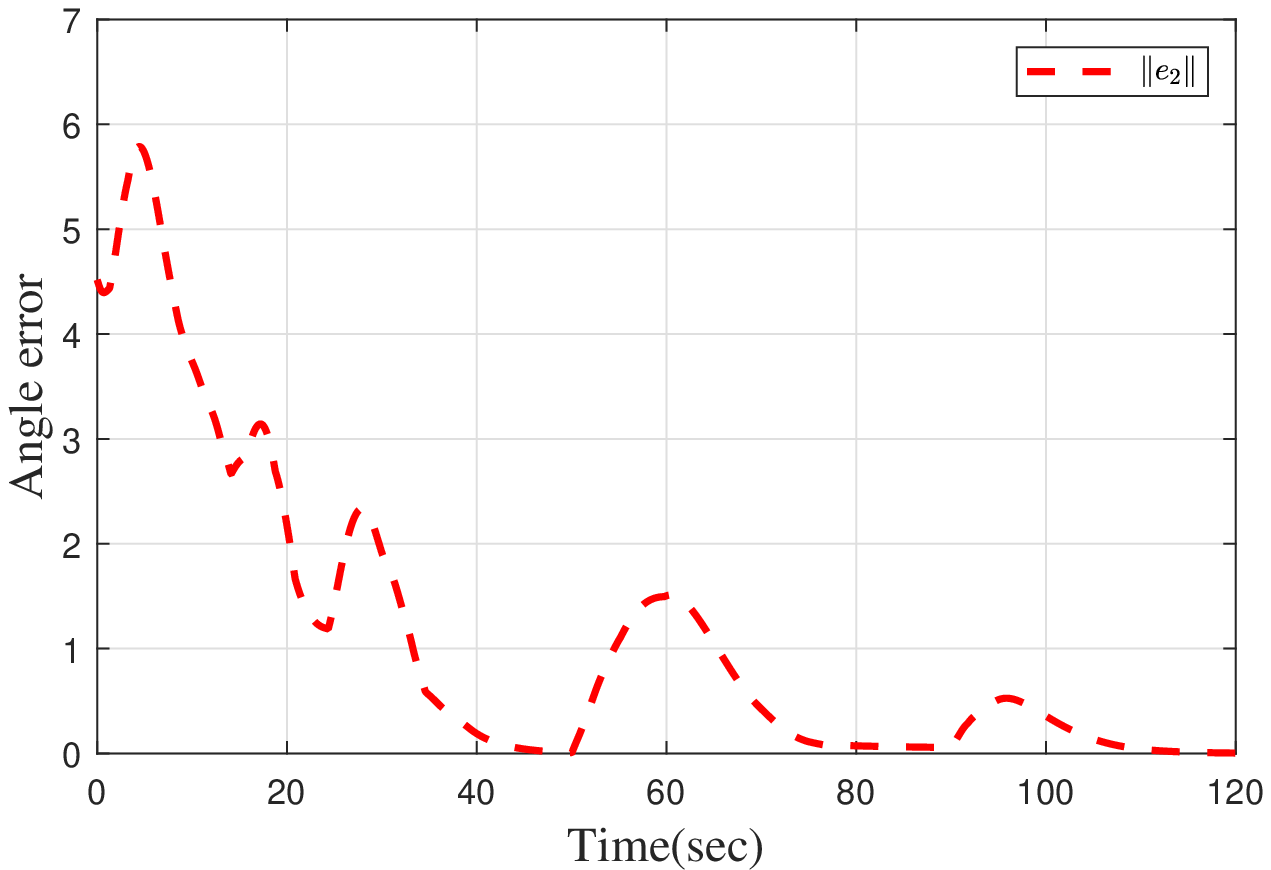}	\label{m error}
	\end{minipage}}
	\centering
	\caption{The formation shape control law (\ref{man}) was implemented on a team of six differential drive robots. (a) initial positions of six differential-drive robots on the Robotarium; (b) the formation at time $t=50s$, $v^{*}_{r}(t)=[0,0.02]^\top$,$\delta^*_{12}(t)=[0.4,0.4]^\top$; (c) the formation at time $t=90s$, $v^{*}_{r}(t)=[0.05,0]^\top$, $\delta^*_{12}(t)=\mathcal{R}(\frac{\pi}{2})[0.4,0.4]^\top$; (d) the formation at time $t=120s$,  $v^{*}_{r}(t)=[0.04,0]^\top$, $\delta^*_{12}(t)=0.7*\mathcal{R}(\frac{\pi}{2})[0.4,0.4]^\top$; (e) evolution of angle errors $||e_2||=\sum_{\alpha_{ijk} \in \mathcal{A}_{\mathcal{G}_f}}\|\alpha_{ijk}-\alpha^*_{ijk}\|$.}\label{figure maneuver}
\end{figure}

This example aims to steer the six differential-drive robots to achieve maneuver tasks cooperatively. The set of desired angle constraints is the same as the one used in the last subsection. The initial positions of six differential-drive robots are set as $p(0)=[-0.4,-1.1,0,0.25,0.5,0.6; -0.35,-0.4,0.025,\\ 0.4,-0.35,0.2]$, which are shown in Fig. \ref{m1}. The posture vectors are randomly chosen from the two-dimensional space. The desired piecewise-constant translational velocities are set as $v^{*}_{r}(t)=[0,0.02]^\top$, $t\in [0,50]$; $v^{*}_{r}(t)=[0.05,0]^\top$, $t\in(50,90]$; $v^{*}_{r}(t)=[0.04,0]^\top$, $t\in(90,120]$; The piecewise-constant relative positions between the leader and the first follower are prescribed as $\delta^*_{12}(t)=[0.4,0.4]^\top$, $t\in [0,50]$; $\delta^*_{12}(t)=\mathcal{R}(-\frac{\pi}{2})[0.4,0.4]^\top$, $t\in (50,90]$; $\delta^*_{12}(t)=0.7*\mathcal{R}(-\frac{\pi}{2}) [0.4,0.4]^\top$, $t\in (90,120]$.
By implementing the control law (\ref{man}), Fig. \ref{figure maneuver} shows the snapshots of the formation in different time intervals.

\section{Acknowledgment}

The authors would like to thank the Robotarium team sincerely for providing a remotely accessible swarm-robotics experiment platform.

\section{CONCLUSION}\label{V}
In this paper, we achieved angle-constrained formation shape stabilization and maneuver control under a directed and non-triangulated sensing graph. The developed distributed controller can be implemented in the local reference frame and ensure global convergence of angle errors. The future work may include angle-constrained formation control with directed sensing graphs in three-dimensional space and angle-constrained formation control with collision avoidance.

\bibliographystyle{automaticreference.bst}
\bibliography{0-reference}
\end{document}